\documentclass[twocolumn,aps,floats,superscriptaddress]{revtex4}
\usepackage{amsmath}
\usepackage{amssymb}
\usepackage{graphicx}

\newcommand{\mr}[1]{{{\mathrm{#1}}}}
\newcommand{\mcal}[1]{{\mathcal{#1}}}
\newcommand{\w}{\omega}
\newcommand{\wt}{\widetilde{\omega}}
\newcommand{\dd}{d_{x^2-y^2}}
\newcommand{\dxy}{d_{xy}}
\newcommand{\Ds}{\Delta_s}
\newcommand{\Dst}{\widetilde{\Delta}_s}
\newcommand{\Dd}{\Delta_d}
\newcommand{\Dxy}{\Delta_{xy}}
\newcommand{\DM}{\Delta_M}
\newcommand{\Dm}{\Delta_m}
\newcommand{\nimp}{n_{\mr{imp}}}
\newcommand{\s}{\sigma}
\newcommand{\R}{{\bf R}}

\begin{document}

\title{Impact of disorder on unconventional superconductors with competing ground states}

\author{Serge Florens and Matthias Vojta}
\affiliation{\mbox{Institut f\"ur Theorie der Kondensierten Materie, Universit\"at
Karlsruhe, 76128 Karlsruhe, Germany}}
\date{\today}

\begin{abstract}
%\vspace{0.5cm}
Non-magnetic impurities are known as strong pair breakers in superconductors with pure
$d$-wave pairing symmetry.
Here we discuss $d$-wave states under the combined influence of impurities and
competing instabilities, such as pairing in a secondary channel as well as
lattice symmetry breaking.
Using the self-consistent T-matrix formalism, we show that disorder
can strongly modify the competition between different pairing states.
For a $d$-wave superconductor in the presence of a subdominant {\em local} attraction,
Anderson's theorem implies that disorder {\it always} generates an $s$-wave component in the
gap at sufficiently low temperature, %and breaks time-reversal symmetry,
even if a pure $\dd$ order parameter characterizes the clean system.
In contrast, disorder is always detrimental to an additional $\dxy$ component.
This qualitative difference suggests that disorder can be used to discriminate among
different mixed-gap structures in high-temperature superconductors.
We also investigate superconducting phases with lattice symmetry breaking
in the form of bond order,
and show that the addition of impurities quickly leads to the {\it restoration} of
translation invariance.
Our results highlight the importance of controlling disorder for the observation
of competing order parameters in cuprates.
\end{abstract}

\maketitle

%%%%%%%%%%%%%%%%%%%%%%%%%%%%%%%%%%%%%%%%%%%%%%%%%%%%%%%%%%%%%%%%%%%%%%%

\section{Introduction}

Although a consensus on the basic $\dd$ symmetry of the superconducting order
parameter in most cuprate materials is now established \cite{TsueiKirtley},
the possibility of further symmetry breaking due to the presence of
competing ground states has been widely discussed, both on theoretical and
experimental grounds.

One of the possible additional order parameters is a secondary superconducting gap of
different symmetry.
In the resulting mixed-gap state, the minimum free energy is usually taken for
complex order parameters like $s+i\dd$ and $\dxy+i\dd$, which implies
the violation of time-reversal invariance (that we will call $\mcal{T}$-breaking),
in addition to the broken U(1) symmetry of the superconducting state.
Experimentally, the observed splitting of the zero-bias conductance peak \cite{Hu94}
of tunneling experiments in the overdoped regime of YBa$_2$Cu$_3$O$_{7-\delta}$ \cite{DaganDeutscher}
has been interpreted as evidence for a $\mcal{T}$-breaking mixed-gap state.
However, the precise nature of this mixed gap is not entirely clear,
as a number of studies reduce to tentative fits of the tunneling spectra using the different gap
structures allowed by symmetry \cite{Sharoni02,Wei98}.
We also point out that the signature of such a mixed gap has remained elusive
in other physical probes.
Surface effects may play a role, as a small secondary gap can be induced by
the proximity to interfaces \cite{MatsumotoShiba,Sigrist98},
although other mechanisms based on competing pairing attractions \cite{SachdevRead,Sangiovanni},
magnetic fields effects \cite{Laughlin98} and presence of magnetic impurities \cite{Balatsky98}
have been proposed.
It is worthwile mentioning that a quantum phase transition, associated with the passage from
the pure $\dd$-wave phase to the mixed superconducting state,
would provide a source of anomalous scattering of the nodal quasiparticles
in the quantum critical regime \cite{VojtaZhangSachdev1,VojtaZhangSachdev2}.
Photoemission measurements in optimally doped Bi$_2$Sr$_2$CaCu$_2$O$_{8+\delta}$ \cite{Valla99}
have indicated the presence of such critical scattering, however, experimental
details have not been fully sorted out.

In electron-doped cuprates the situation appears to be different, as
several experiments on Pr$_{2-x}$Ce$_{x}$CuO${_4}$ indicated a rather sharp transition
from a $\dd$-wave to a $s$-wave state as a function of
doping \cite{Biswas02,Skinta02}.
Therefore a mixed-gap phase could at best be present in a narrow region of the phase diagram.

Of course, in addition other symmetries could be broken in those materials, such as the lattice
symmetry (denoted $\mcal{C}$-breaking in the following).
Indeed, theoretical studies of models for doped Mott insulators
have demonstrated the possibility of several different ordering patterns,
such as bond (or spin-Peierls) order \cite{SachdevRead}, stripes
\cite{zagu,VojtaZhangSachdev3}, and staggered flux states \cite{AffMars}.
On the experimental side, neutron scattering measurements revealed stripe formation
in La$_{2-x}$Sr$_x$CuO$_4$ \cite{tranquada}.
More recently, signatures of charge order were observed by tunneling experiments in
the superconducting phase of Bi$_2$Sr$_2$CaCu$_2$O$_{8+\delta}$
\cite{charge_order,yazdani,kapitul},
giving further evidence that competing orders may be generic to these materials.

In this paper, we argue that disorder plays a significant role for
competing instabilities in unconventional superconductors,
by possibly suppressing or enhancing the tendency for the system to form complex
symmetry breaking patterns. A basic remark is that non-magnetic impurities
are known to strongly suppress the pure $\dd$-wave state, while
usual $s$-wave superconductors are extremely robust to the addition of static disorder,
according to the famous Anderson's theorem \cite{Anderson_theorem}.
This has immediate consequences for the competition of pairing states with different symmetry.
First, a putative phase transition such as $\dd\rightarrow s+i\dd$ could be tuned
by varying the disorder strength (e.g. using irradiation or addition of impurities).
% providing an experimental handle on such an intriguing scenario concerning hole doped cuprates.
Second, we can expect that order parameters of $s+i\dd$ and $\dxy+i\dd$ types should behave
quite differently to the addition of non-magnetic impurities, and this effect (which is
discussed in more detail below) could be used to experimentally discriminate among
the several gap structures discussed for cuprate and other unconventional
superconductors.

To gain some physical understanding, the first part of this paper is devoted to
the simplest theory of disorder in superconductors with a two-component gap.
We argue that the BCS formalism together with a self-consistent T-matrix
approximation \cite{HirschfeldWoelfle} is enough to capture the essential features of
the problem, although quantum critical dynamics~\cite{VojtaZhangSachdev1},
localization phenomena~\cite{HirschfeldAtkinson} and non-uniform pairing
patterns~\cite{Ghosal_s_wave} are not described by our calculation.
%% NEW
In particular, some of our results obtained in the strong disorder regime
should be taken with some caution due to the development of strongly inhomogeneous
regions~\cite{Ghosal_s_wave}. 
Nevertheless, we will be able to determine the phase diagram as a function of the impurity
concentration and of the ratio of two competing pairing attractions, which can
be expected to be qualitatively correct.
Previous studies of this problem \cite{XuTing,GolMaz} were devoted
to the calculation of the critical temperature in the Ginzburg-Landau formalism, and to our
knowledge the solution of the BCS problem with disorder for $s+i\dd$
and $\dxy+i\dd$ superconductors was not investigated in detail.
We shall demonstrate that disorder generates and stabilizes an $s$-wave part in the $s+i\dd$ situation.
This result is due to the combination of two remarkable properties of superconductors: the
insensitivity of the $s$-wave gap to disorder (Anderson's theorem) and the presence of a
finite density of states in a disordered $\dd$-wave state, leading inevitably
to a secondary BCS instability in the $s$-channel at low temperature
(a similar result was discussed in \cite{MeyerGornyiAltland}, but here we are able to provide
a more transparent physical picture).
Because Anderson's theorem does not hold for a $\dxy+i\dd$ superconductor,
the effect of disorder in this case is quite similar to the
pure $\dd$-wave case: superconductivity is rapidly destroyed, and no emergent order parameter
can follow from impurity addition. Because of these two different situations, we
propose that disorder can be used in practice to distinguish the two
debated scenarios in overdoped YBa$_2$Cu$_3$O$_{7-\delta}$, and hope to stimulate new
experimental studies.

The second part of the paper concerns the sensitivity to disorder of superconductors
with spontaneous lattice symmetry breaking, motivated both by stripes physics
and possible charge ordering in cuprates. For simplicity, and because of the local
nature of the T-matrix approximation, we restrict our attention to superconductors
with additional bond order, and demonstrate that the non-magnetic impurities have a
general tendency to restore in an extremely fast manner the global translation
symmetry.%, at least regarding the long-range correlations.

The remainder of the paper is organized as follows:
In Sec.~\ref{sec2} we concentrate on
translation-invariant dirty superconductors with mixed gap symmetry,
and develop the basic formalism, referring the reader to Appendix~\ref{app1}
for technical details.
We investigate in turn the $s+i\dd$ and $\dxy+i\dd$ cases,
discussing the evolution of the superconducting gap as a function of the
impurity concentration and of the ratio of attractive interactions in the two
channels.
In Sec.~\ref{sec3}, we study the more general problem of the influence of
disorder on superconductors that combine U(1) and lattice symmetry breaking.
The complete derivation of the T-matrix formalism is given in Appendix~\ref{app2}.
A general discussion of disorder effects in cuprates is given in Sec. IV,
together with a summary which will close the paper.
Most of the numerical calculations are performed assuming strong impurities,
i.e., unitary-limit scatterers. However, we expect the qualitative results to be
insensitive to details of the disorder potential.

%%%%%%%%%%%%%%%%%%%%%%%%%%%%%%%%%%%%%%%%%%%%%%%%%%%%%%%%%%%%%%%%%%%%%%%

\section{$\mcal{T}$-breaking in disordered superconductors}
\label{sec2}

\subsection{General formalism}

In this Section we will investigate the instabilities of a two-component superconductor
using the weak-coupling BCS theory, owing to the fact that strong-coupling effects
would only modify quantitatively the results, according to previous results
in the Eliashberg framework \cite{ChubuJyont}.
However, the quantum-critical dynamics near the phase transition between
states of different pairing symmetry \cite{VojtaZhangSachdev1,VojtaZhangSachdev2}
cannot be accounted for in our formalism,
as fluctuations of the order parameter are discarded in the mean-field approximation.

Introducing a general pairing attaction $V_{kk'}<0$ in the singlet channel, which
corresponds to the Hamiltonian
\begin{equation}
H = \sum_{k\s} \epsilon_k \psi_{k\s}^\dagger \psi_{k\s}^{\phantom{\dagger}}
+ \sum_{kk'} V_{kk'} \psi_{k\uparrow}^\dagger \psi_{-k\downarrow}^\dagger
\psi_{-k'\downarrow}^{\phantom{\dagger}} \psi_{k'\uparrow}^{\phantom{\dagger}} \,,
\end{equation}
the complex superconducting gap $\Delta_k'+i\Delta_k''$ obeys the BCS equation,
\begin{equation}
\label{BCS}
\Delta_k'+i\Delta_k'' = - \sum_{k'} V_{kk'} \frac{1}{\beta} \sum_n \frac{1}{2}
\mr{Tr} \big[ (\tau^1+i\tau^2) \widehat{g}(k',i\w_n) \big] ,
\end{equation}
where $\beta$ is the inverse temperature and $\w_n = (2n+1)\pi/\beta$ the Matsubara
frequency.
In the following, we will assign the two real gap functions $\Delta_k'$, $\Delta_k''$
to two different representations of the lattice symmetry group, keeping the
global phase of the order parameter fixed.
Further, $\widehat{g}(k,i\omega_n)$ is the single-particle Green's function of the
Nambu spinor $\Psi_k = (\psi_{k\uparrow}, \psi_{-k\downarrow}^{\dagger})$.
It can be decomposed as $\widehat{g} \equiv \sum_\mr{i} g_\mr{i} \tau^\mr{i}$, with $\tau^\mr{i}$,
$\mr{i}=1,2,3$ being the Pauli matrices, and $\tau^0$ is the identity matrix.
The Green's function is determined by Dyson's equation:
\begin{eqnarray}
\nonumber
\widehat{g}(k,i\w_n) &=& \frac{\tau^0}
{i\w_n\tau^0 - \Delta_k'\tau^1 - \Delta_k''\tau^2
-\epsilon_k\tau^3 - \widehat{\Sigma}(k,i\w_n)} \\
\label{Dyson}
&=& - \frac{\;\;i\wt\tau^0+\widetilde{\Delta}_k'\tau^1+\widetilde{\Delta}_k''\tau^2
+\widetilde{\epsilon}_k\tau^3}
{\wt^2+\widetilde{\Delta}_k'^2+\widetilde{\Delta}_k''^2+\widetilde{\epsilon}_k^2}
\end{eqnarray}
where $\epsilon_k$ is the quasiparticle dispersion and
the contribution of impurity scattering to the electronic self-energy is
$\widehat{\Sigma}(k,i\w_n)\equiv\sum_\mr{i}\Sigma_\mr{i}(k,i\w_n)\tau^\mr{i}$.
We have precedently introduced the frequency-dependent renormalized quantities:
\begin{eqnarray}
\label{wt}
i\wt &=& i\w_n - \Sigma_0(i\w_n)\,,\\
\widetilde{\Delta}_k' &=& \Delta_k' + \Sigma_1(i\w_n)\,,\\
\widetilde{\Delta}_k'' &=& \Delta_k'' + \Sigma_2(i\w_n)\,, \label{wt2}\\
\widetilde{\epsilon}_k &=& \epsilon_k + \Sigma_3(i\w_n)\, .
\end{eqnarray}

We will determine $\widehat{\Sigma}$ in the self-consistent T-matrix
approximation, which employs the exact solution of the single impurity
problem.
If we note by $v_{kk'}$ the static potential generated by a given impurity,
the T-matrix obeys the Lippmann-Schwinger equation \cite{HirschfeldWoelfle}:
\begin{equation}
\widehat{T}(k,k',i\w_n) = v_{kk'}\tau^3 + \sum_{k''} v_{kk''}\tau^3
\widehat{g}(k'',i\w_n) \widehat{T}(k'',k',i\w_n) .
\end{equation}
In the case of a finite impurity concentration, $\nimp$, the
T-matrix approximation amounts to the neglect of interference
effects between different impurities.
The main limitation is that localization effects \cite{Lee93} are ignored,
which may be important in low-dimensional systems
(see \cite{HirschfeldAtkinson} and references therein);
also, the occurence of inhomogeneous states due to disorder cannot be
accounted for \cite{Ghosal_s_wave}. Within the T-matrix approach,
disorder averaging effectively restores translation invariance,
and the disorder-induced self-energy is given by the relation:
\begin{equation}
\widehat{\Sigma}(k,i\w_n) = \nimp \widehat{T}(k,k,i\w_n) .
\end{equation}
For simplicity, we will assume a local potential, $v_{kk'}=v$, so that the
solution is easily found:
\begin{equation}
\label{Sigma}
\widehat{\Sigma}(k,i\w_n) = \nimp
\frac{G_0\tau^0-G_1\tau^1-G_2\tau^2+(v^{-1}-G_3)\tau^3}
{-(G_0)^2+(G_1)^2+(G_2)^2+(G_3-v^{-1})^2}
\end{equation}
where we have introduced the local Green's function:
\begin{equation}
\label{Gloc}
G_\mr{i}(i\w_n) \equiv \sum_k g_\mr{i}(k,i\w_n) .
\end{equation}
Equations~(\ref{BCS}),(\ref{Dyson}),(\ref{Sigma}),(\ref{Gloc}) constitute the
starting point of our analysis and must be solved in a fully self-consistent manner.
In the following two paragraphs, we will specialize to the $s+i\dd$ and
$\dxy+i\dd$ cases respectively, which will allow to carry out further the analysis.

\subsection{Disorder in a $s+i\dd$ superconductor}
\label{sec:sid}
We consider here an attractive pairing potential in both $s$ and $\dd$-wave
channels
\begin{eqnarray}
\label{V_sid}
V_{kk'} &=& V_s +
V_d \frac{k_x^2-k_y^2}{k_x^2+k_y^2}\;\frac{k_x'^2-k_y'^2}{k_x'^2+k_y'^2}\\
\nonumber
&=& V_s + V_d \cos(2\phi) \cos(2\phi')
\end{eqnarray}
where $\phi$ is the angle specifying the direction of $k$ in polar coordinates.
This results in $s+i\dd$ superconducting order:
\begin{equation}
\Delta_k = \Ds + i \Dd \cos(2\phi) .
\end{equation}
Ordered states with a relative phase between the two components different from $\pi/2$,
such as $s+\dd$, are easily shown to have a higher free energy \cite{ChubuJyont}, and will
not be considered here.
Here and in the following, we employ the continuum limit,
which also corresponds to the infinite bandwidth limit for the electronic
dispersion.
Particle-hole symmetry then implies $G_3=0$, whereas
the fact that the $d$-wave part of the gap averages to zero on the Fermi surface
gives $G_2=0$.
Performing the $k$ summation in~(\ref{Gloc}) we find the Green's functions:
\begin{eqnarray}
\label{G0_sid}
G_0(i\w_n) &=& - \frac{i\wt \; 2N_0}{\sqrt{\Dd^2+\wt^2+\Dst^2}} F(x) \,,\\
G_1(i\w_n) &=& - \frac{\Dst \; 2N_0}{\sqrt{\Dd^2+\wt^2+\Dst^2}} F(x)
\end{eqnarray}
where $N_0$ is the fermionic density of states,
$F(x)$ is the complete elliptic integral of the first kind \cite{NumRecipes},
and we defined $x \equiv \big[\Dd^2/(\Dd^2+\wt^2+\Dst^2)\big]^{1/2}$.
Using the self-energy~(\ref{Sigma}), we find that the renormalized frequency~(\ref{wt})
is determined by the relation:
\begin{equation}
\label{wt_sid}
\wt = \w_n +
\frac{\wt \; \nimp \; 2 N_0 v^2 \; \frac{F(x)}{\sqrt{\Dd^2+\wt^2+\Dst^2}}}
{1+(2N_0v)^2 (\wt^2+\Dst^2)\frac{F(x)^2}{\Dd^2+\wt^2+\Dst^2}} .
\end{equation}
The renormalized $s$-wave gap $\Dst = \Ds + \Sigma_1(i\w_n)$ is given by an
equation similar to~(\ref{wt_sid}), which we can write in the following way:
\begin{equation}
\label{crucial}
\Dst = \Ds \frac{\wt}{\w_n} .
\end{equation}
This important relation will be central to the later physical discussion. Finally,
we note that $G_2=0$ implies $\Sigma_2=0$, so that the $d$-wave gap does not
acquire a self-energy correction in Eq.~(\ref{wt2}).

The momentum summation in the gap equation~(\ref{BCS}) can be performed similarly,
and leads to
\begin{equation}
\label{gap1_sid}
\Ds = -\frac{1}{\beta} \sum_n^{\;\;\;\;\;\;\; \prime}
\frac{2N_0 \; V_s \; \Dst}{\sqrt{\Dd^2+\wt^2+\Dst^2}} F(x)
\end{equation}
in the $s$-wave channel,
while the $\dd$ gap is determined by:
\begin{eqnarray}
\nonumber
\Dd &=& -\frac{1}{\beta} \sum_n^{\;\;\;\;\;\;\; \prime}
2N_0 \; V_d \; \Dd \frac{\sqrt{\Dd^2+\wt^2+\Dst^2}}{\Dd^2}
\bigg[ E(x) \\
\label{gap2_sid}
& &- \frac{\wt^2+\Dst^2}{\Dd^2+\wt^2+\Dst^2} F(x) \bigg]
\end{eqnarray}
where $E(x)$ is complete elliptic integral of the second kind \cite{NumRecipes}.

The primed summation indicates that a frequency cutoff $\w_c$ needs to be
introduced (see Appendix~\ref{app1}).
We emphasize that this
procedure is crucial in order to preserve Anderson's theorem in a superconductor
with non-retarded interaction such as~(\ref{V_sid}). Indeed, the use of a finite
energy bandwidth together with an unbounded frequency sum would violate this basic
result \cite{KimOverhauser,FayAppel}.
Alternatively, one could employ a retarded pairing interaction from the outset,
without restricting the frequency summation, however, this
procedure would significantly complicate the analysis without modifying
the results.

\subsection{Disorder in a $\dxy+i\dd$ superconductor}

We now consider an attractive pairing potential in both $\dxy$ and $\dd$-wave
channels:
\begin{eqnarray}
V_{kk'} &=& V_{xy} \frac{2 k_x k_y}{2 k_x k_y} \; \frac{2 k_x' k_y'}{2k_x' k_y'}
+ V_d \frac{k_x^2-k_y^2}{k_x^2+k_y^2} \; \frac{k_x'^2-k_y'^2}{k_x'^2+k_y'^2}\\
\nonumber
&=& V_{xy} \sin(2\phi) \sin(2\phi') + V_d \cos(2\phi) \cos(2\phi') .
\end{eqnarray}
This leads to a $\dxy+i\dd$ superconducting order:
\begin{equation}
\Delta_k = \Dxy \sin(2\phi) + i \Dd \cos(2\phi) .
\end{equation}
Arguments discussed previously give $G_1=G_2=G_3=0$, so that both $\Dxy$ and
$\Dd$ are not renormalized. The only non-zero local propagator is:
\begin{equation}
\label{G0_did}
G_0(i\w_n) = - \frac{i\wt \; 2N_0}{\sqrt{\DM^2+\wt^2}} F(y)
\end{equation}
where $y \equiv \big[(\DM^2-\Dm^2)/(\DM^2+\wt^2)\big]^{-1/2}$, with the definitions
$\DM \equiv \mr{Max}(\Dxy,\Dd)$ and $\Dm \equiv \mr{Min}(\Dxy,\Dd)$.
There is therefore a single equation to solve for the renormalized frequency:
\begin{equation}
\label{wt_did}
\wt = \w_n +
\frac{\wt \; \nimp \; 2 N_0 v^2 \; \frac{F(y)}{\sqrt{\DM+\wt^2}}}
{1+(2N_0v)^2 \wt^2\frac{F(y)^2}{\DM^2+\wt^2}} .
\end{equation}

Setting $V_M \equiv \mr{Max}(V_{xy},V_d)$ and
$V_m \equiv \mr{Min}(V_{xy},V_d)$, we finally express the two gap equations:
\begin{eqnarray}
\nonumber
\Dm &=& -\frac{1}{\beta} \sum_n^{\;\;\;\;\;\;\; \prime}
2N_0 \; V_m \; \Dm \frac{\sqrt{\DM^2+\wt^2}}{\DM^2-\Dm^2}
\bigg[F(y) - E(y)\bigg] \\
\label{gap1_did}\\
\nonumber
\DM &=& -\frac{1}{\beta} \sum_n^{\;\;\;\;\;\;\; \prime}
2N_0 \; V_M \; \DM \frac{\sqrt{\DM^2+\wt^2}}{\DM^2-\Dm^2}
\bigg[E(y) \\
\label{gap2_did}
&& - \frac{\wt^2+\Dm^2}{\DM^2+\wt^2} F(y)\bigg]
\end{eqnarray}

We now perform the analytical and numerical analysis of the previous BCS+T-matrix
equations, considering in turn the $s+i\dd$ and $\dxy+i\dd$ cases.

\subsection{Results for the $s+i\dd$ case}

We start by
reviewing the occurence of the quantum phase transition in a clean
superconductor with two competing pairing channels. Then we study the fate of
the critical point in the presence of disorder. Finally we present the phase
diagram obtained from the numerical solution.

\subsubsection{Clean limit}

It is useful to recall the nature of the ground state in a superconductor
with competing $s$ and $d$-wave instabilities. Let us first consider the
possible occurence of an $s$-wave instability in a clean well-formed $\dd$-wave
superconductor. Basically, the BCS singularity is cut off by the presence of the
$\dd$-wave gap, as shown by the fact that the propagator behaves as $(\Dd^2+\w^2)^{-1/2}$ in
the gap equation~(\ref{gap1_sid}). Putting $\Ds$ to zero in this equation, one
sees that a {\it finite} critical interaction $V_s$ is necessary to trigger the
$s$-wave instability, in contrary to the situation in a normal metal. From the same
argument, the $s$-wave state is stable at small $V_d$ up to a critical value
of $V_d$ where the additional $d$-wave order parameter develops. It is straightforward
to establish that there exists a finite range domain in which both orders
coexist at zero temperature (see \cite{ChubuJyont} and Appendix~\ref{app1}):
\begin{equation}
\label{window_sid}
\frac{|V_d|}{2+N_0 |V_d|/2} < |V_s| < \frac{|V_d|}{2} .
\end{equation}
For instance, the quantum critical point associated to the transition
$\dd \rightarrow s+i\dd$ occurs when couplings obey $|V_d|/(2+N_0 |V_d|/2) = |V_s|$
and can be described by a field theory involving the coupling of nodal quasiparticles
to Ising fluctuations associated to the additional $s$-wave order
\cite{VojtaZhangSachdev1,VojtaZhangSachdev2}. -- Clearly, this description
lies beyond the scope of the present mean-field analysis.

\subsubsection{$s$-wave instability in a dirty $\dd$-wave background}

Let us start with an important remark on the effect of disorder in $s$-wave
superconductor ($V_d=0$). The local propagator~(\ref{G0_sid}) reads
\begin{equation}
G_0(i\w_n) = - \frac{i\wt \; \pi N_0}{\sqrt{\wt^2+\Dst^2}} =
- \frac{i\w \; \pi N_0}{\sqrt{\w^2+\Ds^2}} \,,
\end{equation}
where we have used the relation~(\ref{crucial}) to get the r.h.s.
expression.
All renormalization factors in $G_0$ drop out, and therefore
the gap equation~(\ref{gap1_sid}) of the pure $s$-wave superconductor is
{\it unchanged} by the presence of disorder, in accordance with Anderson's
statement.
For the pure $d$-wave case, no such cancellation occurs,
and the gap is rapidly suppressed when a few percent of impurities are added.
We note that, in reality, Anderson's theorem holds to order $\nimp$
\cite{KimOverhauser}, so that the strict insensitivity of $\Ds$ to disorder here
is actually an artefact of the T-matrix approximation.

We now consider a $d$-wave superconductor ($V_s=0$) containing a small amount
of disorder (such that the gap $\Dd$ is not completely driven to zero).
From~(\ref{wt_sid}), assuming unitary scattering ($v\!=\!\infty$) but small impurity
concentration ($\nimp \ll 1$), we find the renormalized frequency
$\wt = \sqrt{\nimp\Dd/(2N_0)}$ at $\w=0^+$ (up to logarithmic corrections coming
from the elliptic function).
This leads to the density of states at the Fermi level:
\begin{equation}
\label{rho}
\rho_0 \equiv -\frac{1}{\pi} \mr{Im} G_0(i0^+) = \sqrt{\frac{\nimp \; 2N_0}{\pi^2\Dd}}
\end{equation}
i.e. a {\it finite} density of states is generated at low energy due to the pair
breaking effect of the non-magnetic impurities \cite{GorkovKalugin,Lee93}.
This result is also valid in the Born limit of small scattering potential $v$,
although $\rho_0$ is exponentially small in $\nimp$ in this case \cite{Lee93}.
Experimentally, scatterers in high-$T_c$ materials appear to be closer to the
unitary limit.

Now, we focus on the $s$-wave attraction and study the related BCS
equation~(\ref{gap1_sid}) in the presence of this disorder-induced density of
states $\rho_0$. By similarity to the BCS instability in a normal metal, we
would naively say that any infinitesimal $V_s$ would lead to a finite $s$-wave
gap $\Ds$. As we will see in the later case of the $\dxy+i\dd$ superconductor, this
argument is true provided Anderson's theorem is fulfilled, which is the case for
$s+i\dd$ superconductors only. Indeed the BCS equation~(\ref{gap1_sid}) can be
re-expressed using again relation~(\ref{crucial}):
\begin{eqnarray}
\nonumber
\Ds &=& - \frac{1}{\beta} \sum_n^{\;\;\;\;\;\;\; \prime}
\frac{2N_0 \; V_s \; \Ds \wt/\w}
{\sqrt{\Dd^2+\wt^2+\Ds^2\wt^2/\w^2}} \,F(x) \\
&=& - \frac{1}{\beta} \sum_n^{\;\;\;\;\;\;\; \prime}
\frac{2N_0 \; V_s \; \Ds \wt}
{\sqrt{\w^2(\Dd^2+\wt^2)+\Ds^2\wt^2}} \,F(x) .
\end{eqnarray}
Because $\wt$ is finite at zero frequency in a dirty $d$-wave superconductor, a $1/\w$
singularity is generated at vanishing $\Ds$ and the weak-coupling BCS instability to the
$s+i\dd$ state occurs, in accordance with our intuitive argument.
Let us summarize this interesting result: {\it due to Anderson's theorem,
an arbitrarily small $s$-wave component of the pairing potential will always generate
an $s$-wave contribution to the order parameter in a disordered $d$-wave superconductor}.
Of course the critical temperature for the onset of the $s$-wave gap is
exponentially small in the disorder-induced density of states (up to
multiplicative factors):
\begin{eqnarray}
\label{eq:Tcs}
T_c^{s-\mr{wave}} &=& \frac{2 N_0 \w_c}{\pi \rho_0} \exp \left(-\frac{1}{\rho_0 |V_s|} \right) \\
\nonumber
&=& \sqrt{\frac{2\Dd N_0}{\nimp}} \w_c \exp \left(-\sqrt{\frac{\pi^2 \Dd}{2 N_0
|V_s| \nimp}} \; \right) .
\end{eqnarray}
In conclusion, this disorder-induced $s$-wave instability could be observed in practice
by increasing the impurity concentration $\nimp$. Note that, while the
exponential term in~(\ref{eq:Tcs}) raises from zero by introducing disorder in the
system, the global prefactor is suppressed by the $\sqrt{\Dd}$ term, and $T_c^{s-\mr{wave}}$
will be maximal for an intermediate value of $\nimp$.

\subsubsection{Phase diagram}

We present now the numerical solution of the self-consistent BCS+T-matrix
equations. In all following calculations, $N_0=1$ is chosen as basic unit, and the
frequency cutoff in the Matsubara sums is $\w_c=0.1$ (such a low value of $\w_c$
allows to obtain realistic gap estimates~\cite{Sangiovanni}). The unitary limit 
($v=\infty$) is also considered here. Importantly, the physics discussed below will 
be qualitatively insensitive to specific values of these parameters.
Considering the mean-field nature of our model we will not try to make quantitative
contact to experimental data.

\begin{figure}
\begin{center}
\includegraphics[width=7.5cm]{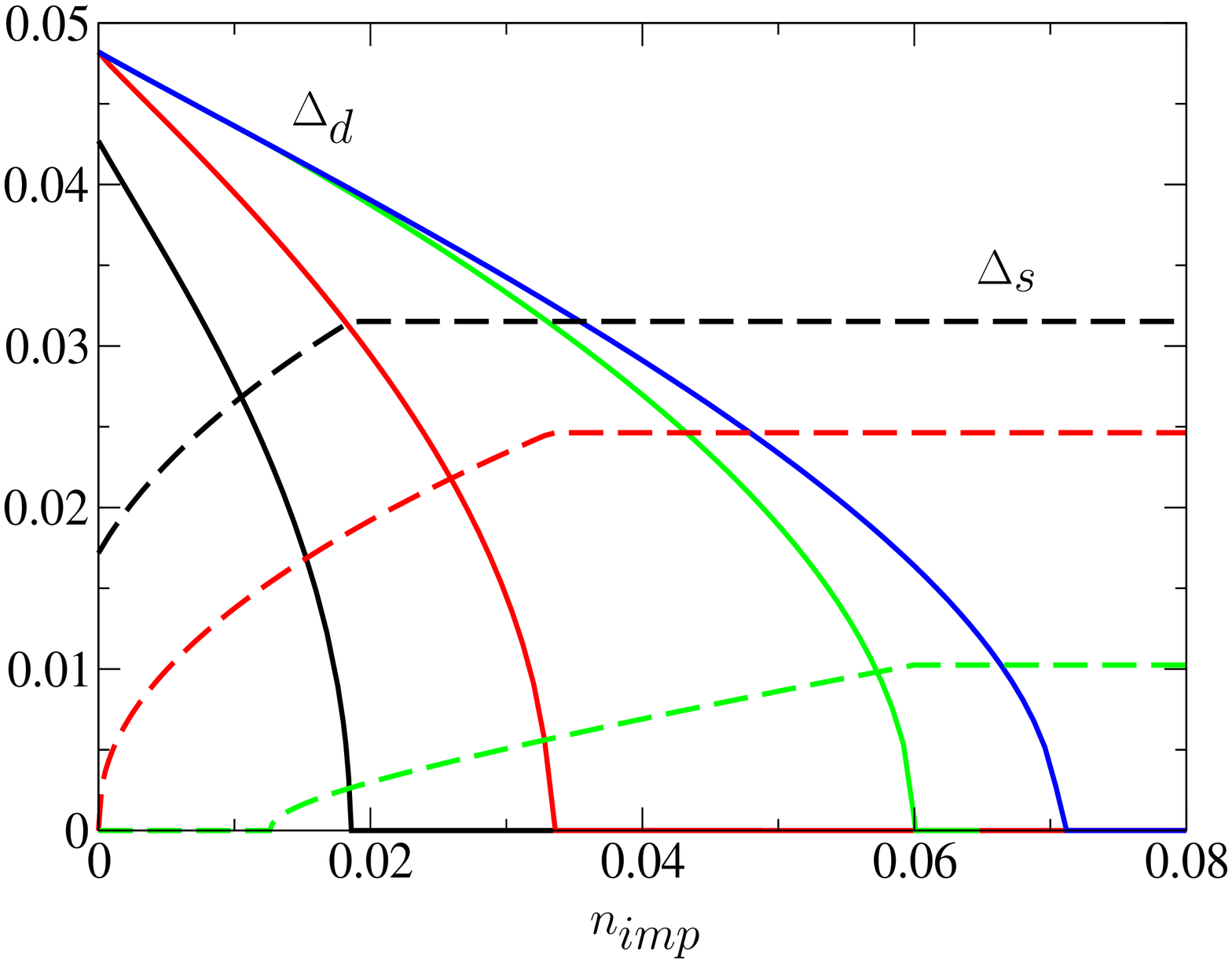}
\end{center}
\caption{(color online).
Superconducting gaps $\Dd$ (continous lines) and $\Ds$ (broken lines)
as a function of $\nimp$ for various values, from left to right, of the $s$-wave pairing
$V_s=-0.45$ (mixed region), $V_s=-0.40$ ($s\rightarrow s+id$ critical point),
$V_s=-0.3$ (clean $d$ phase with $s$ component induced by disorder) and $V_s=-0.1$
(pure $d$ phase). Here $V_d=-1.0$ is fixed and the inverse temperature is taken
as $\beta=1000$.}
\label{sweep_sid}
\end{figure}

\begin{figure}
\begin{center}
\includegraphics[width=7.5cm]{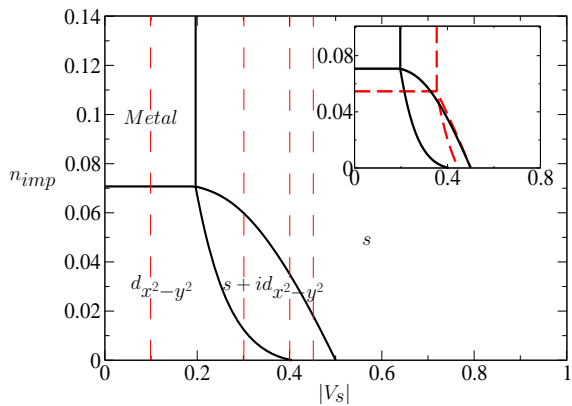}
\end{center}
\caption{(color online). Phase diagram ($|V_s|$,$\nimp$) of the dirty $s+i\dd$
superconductor for $\beta=1000$.
Thin dashed lines refer to the gaps plotted in Fig.~\protect\ref{sweep_sid}.
The inset illustrates the evolution of the phase boudaries with temperature, here
shown for $\beta=100$ (broken lines) and $\beta=1000$ (continuous lines).
}
\label{phase_sid}
\end{figure}

\begin{figure}
\begin{center}
\includegraphics[width=8.0cm]{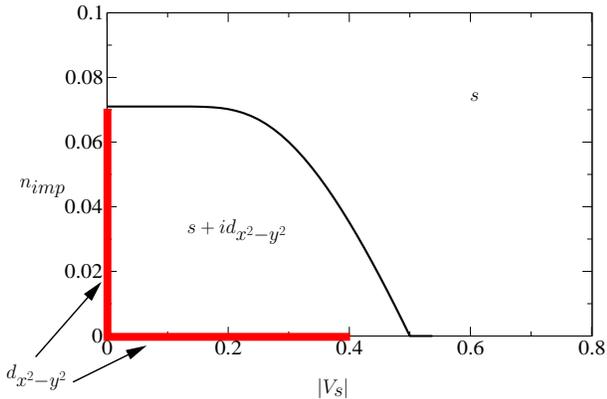}
\end{center}
\caption{(color online). Zero-temperature phase diagram ($|V_s|$,$\nimp$) in the
dirty $s+i\dd$ case. The thick line stands for the limited pure $\dd$ region.
Compare with the finite temperature plot displayed in the inset of
Fig.~\ref{phase_sid}.}
\label{zeroT_sid}
\end{figure}

Fig.~\ref{sweep_sid} illustrates our main result by plotting the evolution of $\Dd$
and $\Ds$ as a function of impurity concentration $\nimp$. We notice that $\Dd$
is generically suppressed by the addition of disorder, whereas $\Ds$ is generated
even if not initially present in the clean system (assuming $V_s \neq 0$). Because
these curves are computed at finite temperature, a critical interaction $V_s$ is
necessary to enter the $s+i\dd$ state, as shown in the related phase diagram,
Fig.~\ref{phase_sid}.
However, when temperature is lowered, the pure $\dd$-wave phase is gradually suppressed
due to the unavoidable $s$-wave instability discussed previously, leading to the interesting
zero-temperature phase diagram shown in Fig.~\ref{zeroT_sid}.

\subsection{Results for the $\dxy+i\dd$ case}
\subsubsection{Clean limit}

In the absence of disorder, $\dxy$ and $\dd$ order parameters compete as in the
previous case, leading to a pure $\dd$-wave state at small $V_{xy}$, a pure
$\dxy$-wave state at large $V_{xy}$ and a mixed region of $\dxy+i\dd$ symmetry
in between (the $\dxy+\dd$ state being unfavorable energetically). This region
is found to be given by the condition (see Appendix~\ref{app1}):
\begin{equation}
\label{window_did}
\frac{|V_d|}{1+N_0 |V_d|/2} < |V_{xy}| < \frac{|V_d|}{1-N_0 |V_d|/2} .
\end{equation}

Of course, neither the pure $\dxy$ phase nor the pure $\dd$ phase obey
Anderson's theorem. This will have important consequences for the following
analysis.

\subsubsection{$\dxy$-wave instability in a dirty $\dd$-wave background}

We now consider the effect of a small pairing amplitude in the $\dxy$ channel,
supposing that the $\dd$ order is present despite the finite amount of
disorder.
Although a finite density of states
$\rho_0$ is present, we easily see that the naive argument predicting a
weak-coupling BCS instability for the $\Dxy$ gap {\it does not hold}.
Indeed the gap equation~(\ref{gap1_did}) in the $\dxy$ channel reads:
\begin{equation}
\Dxy = -\frac{1}{\beta} \sum_n^{\;\;\;\;\;\;\; \prime}
2N_0 \; V_{xy} \; \Dxy \frac{\sqrt{\Dd^2+\wt^2}}{\Dd^2-\Dxy^2}
\bigg[F(y) - E(y)\bigg] .
\end{equation}
The integral is completely regular at low frequency, and $\Dxy$ can be put to
zero, leading to a critical interaction $V_{xy}$ to generate the $\dxy$-wave gap,
similarly to the clean case. The physics is therefore very different from the
$s+i\dd$ situation, and is due to the absence of Anderson's theorem in
$\dxy$-wave superconductors.

\subsubsection{Phase diagram}

\begin{figure}
\begin{center}
\includegraphics[width=7.5cm]{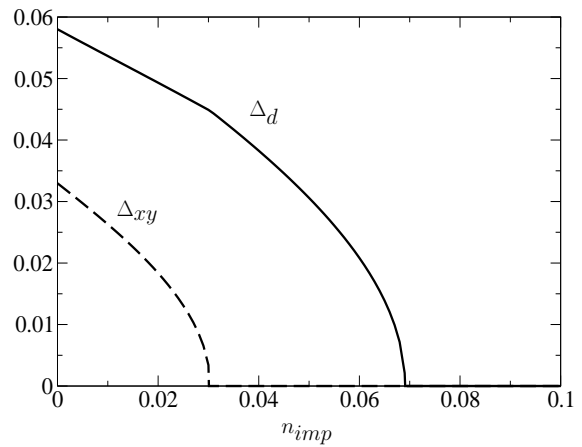}
\end{center}
\caption{Superconducting gaps $\Dd$ (continous line) and $\Dxy$ (broken line)
as a function of $\nimp$ for the $\dxy$-wave pairing $V_{xy}=-0.47$ (mixed phase in
the clean limit), with $V_d=-0.5$ and $\beta=100$.
}
\label{sweep_did}
\end{figure}

The previous discussion implies that the results for the $\dxy+i\dd$ case are
markedly different from the one obtained in the $s+i\dd$ case. Indeed, the
numerical solution of the mean-field equations show that, if the $\dxy$-wave gap is not
formed in the clean limit, it cannot be generated by adding impurities. By the
same token, in the mixed region, both order parameters $\Dd$ and $\Dxy$
are gradually suppressed. This is shown in Fig.~\ref{sweep_did}.

\begin{figure}
\begin{center}
\includegraphics[width=7.5cm]{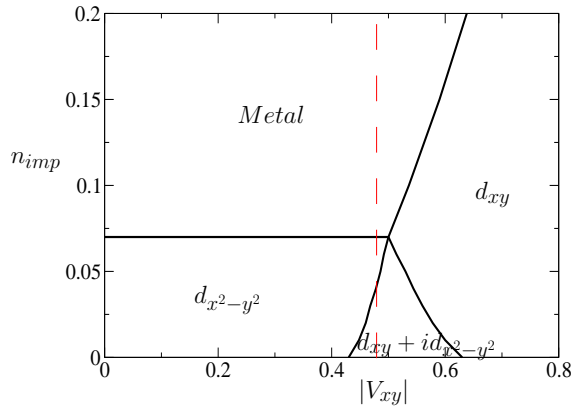}
\end{center}
\caption{Phase diagram ($|V_{xy}|$,$\nimp$) of the dirty $\dxy+i\dd$ superconductor,
with $V_d=-0.5$ and $\beta=100$.
The thin dashed line refers to the gaps drawn in Fig.~\ref{sweep_did}.}
\label{phase_did}
\end{figure}

The phase diagram illustrates this behavior, see Fig.~\ref{phase_did}.
We stress that the zero-temperature phase diagram is quite similar to this
picture, i.e., the pure $\dd$-wave region remains extended, in contrast to the
$s+i\dd$ case.

%%%%%%%%%%%%%%%%%%%%%%%%%%%%%%%%%%%%%%%%%%%%%%%%%%%%%%%%%%%%%%%%%%%%%%%

\section{$\mcal{C}$-breaking in disordered superconductors}
\label{sec3}

In contrast to conventional superconductors, cuprates derive from their
Mott insulating parent compounds, and the strong-correlation physics
of doped Mott insulators is responsible for many of the intriguing
phenomena observed in these high-$T_c$ materials.
Theoretical studies of microscopic models, such as the $t-J$ and Hubbard
models on a two-dimensional square lattice, have shown that a variety
of ordering tendencies compete, and that the actual ground state may
depend sensitively on microscopic details.
Experimentally, some compounds shown evidence for lattice symmetry
breaking in form of charge and/or spin order, possibly coexisting with
superconductivity at low temperatures.

\begin{figure}
\begin{center}
\includegraphics[width=7.0cm]{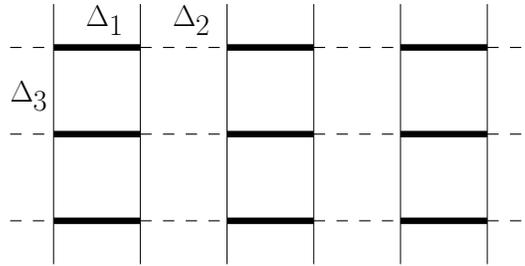}
\end{center}
\caption{Gap structure of a bond-ordered superconductor (dubbed spin Peierls
in the following). In general, gaps take complex values, so that time
reversal symmetry is also broken.}
\label{bond}
\end{figure}

In this section, we shall consider the competition between superconductivity and
additional order associated with broken translation symmetry
under the influence of disorder.
Many candidates for the additional order could be chosen \cite{VojtaZhangSachdev3}, but we will concentrate
here on the simplest case of additional bond (or spin-Peierls) order,
where bond strengths are modulated as shown in Fig.~\ref{bond}.
Such a state obeys a two-site unit cell, but up to a rotation all sites are
equivalent.
This justifies the application of the T-matrix approximation,
based on the exact solution of a {\it single} defect,
where the use of a single self energy is sufficient.
In contrast, for more complicated forms of order, such as stripe states
with varying site charge density, the T-matrix approach would clearly
be inappropriate,
and this is one of the reasons to restrict our attention to bond order.

\subsection{Formalism}

We start by developing the
necessary formalism in the absence of disorder, then compute the T-matrix in
such a non-translation-invariant superconducting background, which will finally lead
us to study how disorder affects the chosen ordered state.

\subsubsection{Clean case}

Contrarily to Sec.~\ref{sec2}, we cannot directly simplify the analysis in the continuum
limit as the lattice symmetry is now explicitely broken. The simplest discrete model
that exhibits the physics we are interested in is given by the $t-J$ Hamiltonian:
\begin{equation}
H_{tJ} = \sum_{\R \R'\s}
\left[ -t_{\R\R'} c_{\R\s}^\dagger c_{\R'\s}^{\phantom{\dagger}} +
J_{\R\R'} \left( {\bf S}_{\R} \cdot {\bf S}_{\R'} -
\frac{{n}_{\R} {n}_{\R'}}{4} \right) \right]
\label{e1}
\end{equation}
where the ${c}_{\R\s}^{\dagger}$ operators prohibit from double occupancies,
the rest of the notation being standard.
Restricting both hopping $t$ and exchange interaction $J$ to nearest-neighbor sites,
and introducing a slave boson $b_{\R}^\dagger$ to implement the non-double occupancy
restriction, with the representation
$c_{\R\s}^\dagger = f_{\R\s}^\dagger b_{\R}^{\phantom{\dagger}}$
and
the constraint $\sum_\s f_{\R\s}^\dagger f_{\R\s}^{\phantom{\dagger}} +
b_{\R}^\dagger b_{\R}^{\phantom{\dagger}} = 1$,
we arrive at
\begin{eqnarray}
\label{Heff}
H & = & -t \sum_{\left<\R \R'\right>\s} b_{\R}^{\phantom{\dagger}} b_{\R'}^\dagger
f_{\R\s}^\dagger f_{\R'\s}^{\phantom{\dagger}}
- \mu \sum_{\R \s} f_{\R\s}^\dagger f_{\R\s}^{\phantom{\dagger}} \\
\nonumber
& - & J \sum_{\left<\R \R'\right> \s s} (-1)^\s f_{\R\s}^\dagger f_{\R'-\s}^\dagger
(-1)^{s} f_{\R'-s}^{\phantom{\dagger}} f_{\R s}^{\phantom{\dagger}}
\end{eqnarray}
where
$\left<\R \R'\right>$ denotes nearest neighbor pairs of lattice sites, and
$\mu$ is the chemical potential enforcing the constraint.
We have organized the interaction
term in such a way that the mean-field treatment in the particle-particle sector
(controlled by a large-$N$ limit in the Sp($N$)
formulation~\cite{SachdevRead,VojtaZhangSachdev3}) is quite straightforward:
\begin{eqnarray}
\label{HMF}
H_{MF} & = & -t_\mr{eff} \sum_{\left<\R \R'\right>\s} f_{\R\s}^\dagger f_{\R'\s}^{\phantom{\dagger}}
- \mu \sum_{\R \s} f_{\R\s}^\dagger f_{\R\s}^{\phantom{\dagger}} \\
\nonumber
& + & \sum_{\left<\R \R'\right> \s} (-1)^\s \Delta_{\R \R'} f_{\R\s}^\dagger f_{\R'-\s}^\dagger
+ h.c.
\end{eqnarray}
with
\begin{eqnarray}
\label{DeltaRR}
\Delta_{\R \R'} & = & -J \sum_s (-1)^{s} \big<f_{\R'-s}^{\phantom{\dagger}}
f_{\R s}^{\phantom{\dagger}}\big>
\end{eqnarray}
This mean-field is equivalent to the pairing part within the BCS formalism
(which can be argued~\cite{SC_flex} to be enough even when pairing is induced by 
the presence of strong correlations).
The slave bosons $b_{\R}$ in (\ref{Heff}) condense at low temperatures, and at the mean-field
level the $b_{\R}$ operators are simply replaced by their condensation amplitude
$b_{\R}^\dagger = \big<b_{\R}^{\dagger}\big> = \sqrt{\delta}$, where $\delta$ is the hole
doping, $\sum_{R\s} f_{\R\s}^\dagger f_{\R\s}^{\phantom{\dagger}} = 1-\delta$.
Thus, the strong Coulomb repulsion in the model (\ref{e1}) is accounted for by
a renormalized hopping $t_\mr{eff} = t \delta$ in (\ref{HMF}).

If we assume the symmetry breaking pattern shown in Fig.~\ref{bond}, we have
two sites per unit cell. On these two sublattices, we can define the
fermions $f_{\R\s}^\dagger \equiv c_{Ij\s}^\dagger$ for
$\R=2I{\bf e}_x+j{\bf e}_y$ and $f_{\R\s}^\dagger \equiv d_{Ij\s}^\dagger$ for
$\R=(2I+1){\bf e}_x+j{\bf e}_y$, where $I$ and $j$ are integers. Generalizing the
Nambu formalism, we introduce a four-component spinor:
\begin{equation}
\Psi^\dagger \equiv
(c_{K_x k_y \uparrow}^\dagger \;\; c_{-K_x -k_y \downarrow}^{\phantom{\dagger}}
\;\; d_{K_x k_y \uparrow}^\dagger \;\; d_{-K_x -k_y \downarrow}^{\phantom{\dagger}})
\end{equation}
with $K_x=2k_x$ the supermomentum associated to $I$ and $k_y$ the usual momentum
associated to $j$. After some manipulations (see Appendix~\ref{app2}), we find the
mean-field Hamiltonian in matrix form $H_{MF} = \sum_{K_x k_y} \Psi^\dagger \hat{h}_k \Psi$,
where:
\begin{widetext}
\begin{align}
\label{Hmatrix}
\hat{h}_k = \left[
\begin{array}{cccc}
-2 t_\mr{eff} \cos k_y -\mu & 2\Delta_3 \cos k_y & -t_\mr{eff}-t_\mr{eff}e^{-iK_x} & \Delta_1+\Delta_2 e^{-iK_x} \\
2\Delta_3^\star \cos k_y & 2 t_\mr{eff}\cos k_y + \mu & \Delta_1^\star+\Delta_2^\star e^{-iK_x} & t_\mr{eff}+t_\mr{eff}e^{-iK_x} \\
-t_\mr{eff}-t_\mr{eff}e^{iK_x} & \Delta_1+\Delta_2 e^{iK_x} & -2 t_\mr{eff}\cos k_y - \mu & 2\Delta_3 \cos k_y \\
\Delta_1^\star+\Delta_2^\star e^{iK_x} & t_\mr{eff}+t_\mr{eff}e^{iK_x} & 2\Delta_3^\star \cos k_y & 2t_\mr{eff}\cos k_y + \mu
\end{array}
\right]
\end{align}
\end{widetext}
The gap equation~(\ref{DeltaRR}) can similarly expressed (see Appendix~\ref{app2})
in terms of Green's functions of the $\Psi$ spinor:
\begin{equation}
\label{dysonMatrix}
\hat{g}^{-1} = i\w_n - \hat{h}_k - \hat{\Sigma}
\end{equation}
where $\hat{\Sigma}$ contains effects due to the disorder distribution, that we will
evaluate now. In principle, $\hat{\Sigma}$ has also contributions due
to fluctuation effects beyond mean-field, which we will not take into account here.

\subsubsection{Calculation of the T-matrix}

The calculation of the T-matrix corresponds to solving the problem of a single
localized impurity at ${\bf R = 0}$ in the superconducting background state.
It is obtained from the solution of the problem $H_{MF} + v \sum_{\s} f_{{\bf 0}\s}^\dagger
f_{{\bf 0} \s}^{\phantom{\dagger}}$, and after some manipulations, we find (see
Appendix~\ref{app2}):
\begin{equation}
\hat{T}(i\w_n) = \frac{1}{G_{11}G_{22} - G_{12}G_{21}}
\left[
\begin{array}{cccc}
-G_{22} & G_{21} & 0 & 0 \\
G_{12} & G_{11} & 0 & 0 \\
0 & 0 & -G_{44} & G_{43} \\
0 & 0 & G_{34} & G_{33}
\end{array}
\right] \\
\label{Tmatrix}
\end{equation}
where $G_{mn}(i\w_n) = \sum_{K_x k_y} g_{mn}(i\w_n,K_x,k_y)$ is the local propagator.
For simplicity, we have assumed the unitary limit $v = +\infty$.

\subsubsection{Self-consistency cycle}

In the presence of a finite but small concentration $\nimp$ of impurities, we can
approximate the self-energy by $\hat{\Sigma} = \nimp \hat{T}$. Self-consistency
must be reached at two levels: first because the T-matrix is
expressed through the propagator by the result~(\ref{Tmatrix}) and Dyson's
equation~(\ref{dysonMatrix}), second because the gaps $\Delta_1, \Delta_2,
\Delta_3$ should obey the BCS equations~(\ref{D1}-\ref{D3}). This whole procedure is
performed numerically, with a discrete momentum mesh and at small finite
temperature.

\subsection{Numerical results}
\subsubsection{Clean case}

\begin{figure}
\begin{center}
\includegraphics[width=7.0cm]{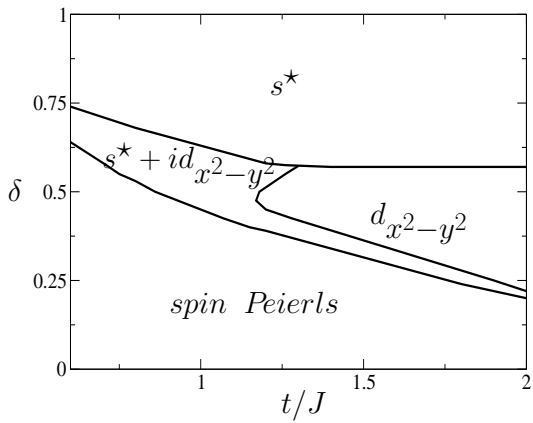}
\end{center}
\caption{Superconducting phase diagram of the clean $t-J$ model for $\beta=100$
(in units of $J=1$), with a two-site unit cell. The transition from $s^\star$
to $\dd$ is discontinuous, whereas all other phase boundaries are second order.
These results are consistent with~\protect\cite{SachdevRead}.}
\label{phase_clean}
\end{figure}

In the absence of disorder, the Hamiltonian~(\ref{Heff}) shows several phases as
a function of the pairing attraction $J$ and doping $\delta$ \cite{SachdevRead,VojtaZhangSachdev3}.
The $\mcal{C}$-broken bond-order phase, which coexists with superconductivity,
occurs typically at low doping, and is characterized by the $\Delta_1,\Delta_2,\Delta_3$ all being different.
Because the effective hopping
$t_\mr{eff} = t \delta$ can become much smaller than $J$ in this regime, this corresponds
to a strong-coupling superconductor.
The bond-ordered state also breaks $\mcal{T}$, as the
three bond variables cannot be made all real by a global gauge transformation.
At intermediate doping, the ground state is uniform, but continues to break
time-reversal symmetry, as the $s^\star+i\dd$ phase happens to be more stable
(if $J$ is large enough only, otherwise the pure $\dd$-wave state with
$\Delta_1=\Delta_2=-\Delta_3$ is found).
Finally, an extended
$s$-wave phase, denoted $s^\star$ and characterized by $\Delta_1=\Delta_2=\Delta_3$,
is encountered at larger doping.
We have checked numerically these known results for the clean case, as shown in
Fig.~\ref{phase_clean}.

\subsubsection{Dirty case}

We investigate now how these phases evolve with the addition of non-magnetic
impurities. For this model calculation, we stress that Anderson's theorem does
not hold, as a finite energy bandwidth is used (see the discussion at the end
of Sec.~\ref{sec:sid}). For this reason, and in contrast to the results
obtained in the first part of the paper, we do not see transitions from $\dd$ to
$s^\star+i\dd$ by increasing the disorder, and rather all superconducting phases
are destroyed in a similar fashion.

\begin{figure}
\begin{center}
\includegraphics[width=7.0cm]{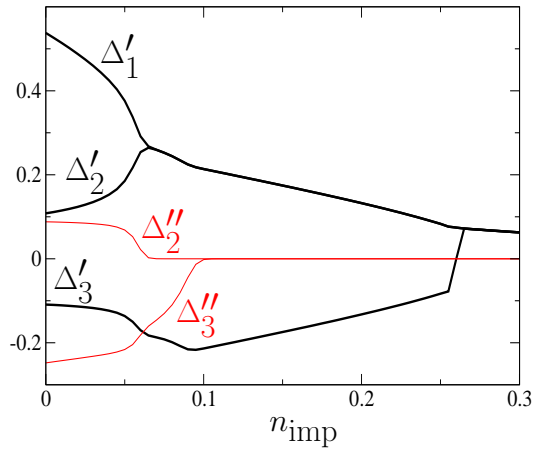}
\end{center}
\caption{(color online). Superconducting gaps as a function of impurity concentration, for
$t/J=1$, $\beta=100$ and $\delta=0.375$. Single (double) primed quantities
denote real (imaginary) parts, respectively. Here we chose a gauge where
$\Delta_1'=0$.
The initial state at $\nimp=0$ is the bond-ordered superconductor described in the text.}
\label{SP_to_s}
\end{figure}

Let us discuss the effect of adding impurities to the bond-ordered phase.
The numerical solution of
the problem demonstrates that the translation symmetry is restored during this
process, i.e. {\it the $\mcal{C}$-broken phase is generically unstable to the
addition of disorder}, as displayed in Fig.~\ref{SP_to_s}.
This result can be expected on the basis that impurities disrupt the perfect
arrangement of superconducting dimers, suppressing their long-range order.
(In principle, it is possible that dimers locally arrange around impurities, resulting
in enhanced {\em local} dimer correlations \cite{Ghosal_s_wave}, but this is not
related to bond long-range order, and is certainly beyond the scope of our mean-field approach.)
Discussing in more detail the evolution of the bond parameters
shown in this figure, we see that the strongly dimerized superconductor persists
for $0<\nimp<0.07$, while being progressively suppressed. We then have a narrow
$s^\star+i\dd$ phase with restored lattice symmetry for $0.07<\nimp<0.10$
(in which $\Delta_1''=\Delta_2''=0$ and $\Delta_1'=\Delta_2'=|\Delta_3'+i\Delta_3''|$).
Finally, a $\dd$-wave state (gaps are real, with $\Delta_1=\Delta_2
=-\Delta_3$) is found at larger impurity concentration, ultimately unstable
towards the $s^\star$ phase (with $\Delta_1=\Delta_2=\Delta_3$) for $\nimp>0.25$ in
a discontinuous manner.

\begin{figure}
\begin{center}
\includegraphics[width=7.0cm]{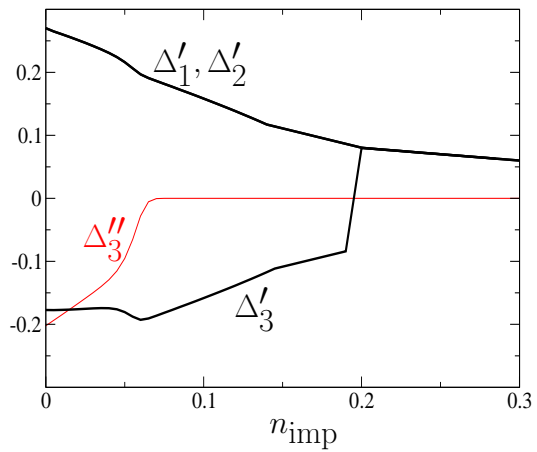}
\end{center}
\caption{(color online). Superconducting gaps as a function of impurity concentration, same
conditions as in Fig.~\ref{SP_to_s}, with $\delta=0.45$ here. The initial state
at $\nimp=0$ is $s^\star+i\dd$.}
\label{sid_to_s}
\end{figure}

Similarly, starting at larger doping from the clean $s^\star+i\dd$
phase, Fig.~\ref{sid_to_s} shows an initial suppression with
$\nimp$ of the $\mcal{T}$-breaking towards the $\dd$-wave phase,
which finally displays a first-order transition into the extended $s^\star$-wave
phase.
In all these calculations, we note that the unrealistic value for the
critical concentration, where all superconductivity ultimately vanishes,
is an artifact of the mean-field approximation:
to stabilize the bond-ordered phase, a very small $t_{\rm eff}$ is required,
placing the superconductor effectively into the strong-coupling limit
which is rather robust against impurities.
Nevertheless, we expect our results to correctly capture the qualitative aspects
of impurity doping.

\begin{figure}
\begin{center}
\includegraphics[width=7.0cm]{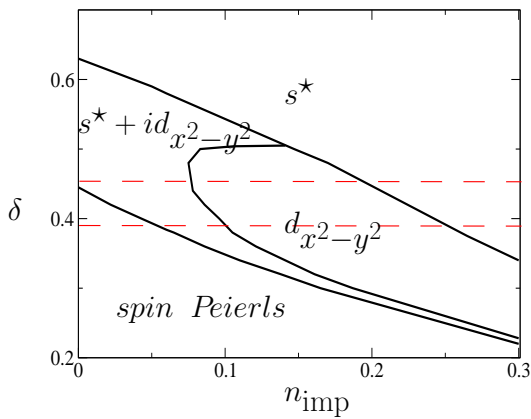}
\end{center}
\caption{Superconducting phase diagram of the dirty $t-J$ model as a function of impurity
concentration $\nimp$ and doping $\delta$, for $t/J=1$ and $\beta=100$.
The two dashed lines refer to the cuts in Figs.~\ref{SP_to_s}
and~\ref{sid_to_s}, respectively.}
\label{phase_dirty}
\end{figure}

The described sequence of phases as a function of impurity concentration can
roughly be guessed from the knowledge of the phase diagram in the clean case.
We summarize our results for the dirty $\mcal{C}$-breaking superconductor in
Fig.~\ref{phase_dirty}.
We have not mapped out the full set of phase boundaries for all possible
parameter combinations, thus for different values of $t/J$, the
various phases might organize themselves in a different way.

\section{Discussion and conclusions}

In this paper, we have investigated the influence of static disorder on
superconductors with competing instabilities. Regarding the possibility of
$\mcal{T}$-breaking states with a mixed gap of $s+i\dd$ and $\dxy+i\dd$
structure, we found important qualitative differences between those two cases
in the process of adding non magnetic impurities. In studying next
a superconductor with bond long-range order (doped spin-Peierls state), we found that
the lattice symmetry is very quickly recovered by addition of disorder, and we
would expect this effect to be generic.

\subsection{Application to high-temperature superconductors}

The finite-temperature phase diagrams displayed in Fig.~\ref{phase_sid}
for the $s+i\dd$ superconductor and Fig.~\ref{phase_did} for the
$\dxy+i\dd$ superconductor show clear {\it qualitative} differences to
the effect of impurity addition. Indeed, the minor $s$-component is enhanced
with respect to the dominant $\dd$-wave background in the first case, whereas
the small $\dxy$ part of the gap is very rapidly suppressed in the second case.
This implies that typical tunelling spectra would evolve in a very different
manner with impurity concentration, depending of the underlying structure of the
superconducting gap: in the $s+i\dd$ case, we expect that the zero-bias
conductance peak splitting should increase, while the opposite behavior
could be seen in the $\dxy+i\dd$ case. It would thus be very interesting to see
such an experimental study in overdoped YBa$_2$Cu$_3$O$_{7-\delta}$, which could
lead to a definite statement on the nature of the mixed superconducting gap
in this compound. We recognize, however, that the practical situation might be
complicated by the fact that magnetic moments can be induced by
non-magnetic scatterers such as Zn \cite{Bobroff99}, which then would strongly
affect the $s$-wave component as well.

A second point we would like to emphasize is that the typical coexistence windows given in
equations~(\ref{window_sid}) and~(\ref{window_did}) are extremely narrow in the
weak-coupling regime $N_0 V_d \ll 1$. Because of their smaller critical temperature, we
believe that electron-doped superconductors as Pr$_{2-x}$Ce$_x$CuO$_4$ fall into this
condition, explaining the absence of clear observation of a mixed phase in tunneling
experiments \cite{Biswas02}.

Let us turn to states with additional breaking of translation symmetry, i.e.,
charge order.
Our calculation nicely shows that the {\em global} symmetry is restored very quickly
upon adding impurities.
Clearly, physics beyond mean-field can lead to the nucleation of {\em local}
symmetry-breaking patterns around impurities due to pinning effects.
Then, in the presence of a sufficient amount of impurity disorder, signatures
of charge ordering will only be visible in local probes, such as STM, but
no sharp superlattice Bragg peaks will be seen, and no thermodynamic phase
transition will occur.
This scenario definitely has similarities with what is seen in certain
cuprate compounds \cite{charge_order,yazdani,kapitul}.

Competing orders are ubiquitous to many exotic superconductors.
Thus we think the theory developed in this paper would likely find applications
in some other context.
Candidate examples are Na$_x$CoO$_2$ \cite{nacoo}, a possible realization of a
doped Mott insulator on the triangular lattice, and
organic superconductors of the BEDT-TTF family \cite{bedtttf}, where
both spin and charge fluctuations seems to play a crucial role for the
pairing mechanism.

\acknowledgments

We thank P. Hirschfeld, X. Wan, and P. W\"olfle for helpful discussions.
This research was supported by the Deutsche Forschungsgemeinschaft
through the Center for Functional Nano\-structures Karlsruhe.

%%%%%%%%%%%%%%%%%%%%%%%%%%%%%%%%%%%%%%%%%%%%%%%%%%%%%%%%%%%%%%%%%%%%%%%

\appendix

\section{Derivation of the gap equation for $\mcal{T}$-breaking
superconductors}
\label{app1}
\subsection{$s+i\dd$ superconductor}

In order to set up clearly the notation, we present how the gap
equations~(\ref{gap1_sid})-(\ref{gap2_sid}) can be obtained from~(\ref{BCS}).
We first transform the k sums into an integral over angle and energy in the infinite bandwidth
limit, allowing Anderson's theorem to be preserved \cite{KimOverhauser,FayAppel}:
\begin{eqnarray}
\nonumber
\Ds &=& - \frac{1}{\beta} \sum_n^{\;\;\;\;\;\;\; \prime} \int_0^{2\pi} \!\!\! \frac{\mr{d}\phi}{2\pi}
\int_{-\infty}^{+\infty} \!\!\!\!\!\!\! \mr{d}\epsilon \frac{N_0 V_s \Dst}
{\wt^2+\Dst^2+\Dd^2\cos^2(2\phi)+\epsilon^2} \\
\nonumber
\Dd &=& - \frac{1}{\beta} \sum_n^{\;\;\;\;\;\;\; \prime} \int_0^{2\pi} \!\!\! \frac{\mr{d}\phi}{2\pi}
\int_{-\infty}^{+\infty} \!\!\!\!\!\!\! \mr{d}\epsilon \frac{N_0 V_d \Dd\cos^2(2\phi)}
{\wt^2+\Dst^2+\Dd^2\cos^2(2\phi)+\epsilon^2} .
\end{eqnarray}
Because of the self-consistency, $\wt$ and $\Dst$ are unknown functions of the
frequency $\w_n$, and the Matsubara sum cannot be performed analytically.
However, the integral over $\epsilon$ is straightforward to do:
\begin{eqnarray}
\label{GAP1_sid}
\!\!\!\!\!\!\! \Ds &=& - \frac{1}{\beta} \sum_n^{\;\;\;\;\;\;\; \prime}
\int_0^{\pi/2} \!\!\!\!\!\!\! \mr{d}\phi
\frac{2N_0 V_s \Dst}{\sqrt{\wt^2+\Dst^2+\Dd^2\cos^2\phi}} \,, \\
\label{GAP2_sid}
\!\!\!\!\!\!\! \Dd &=& - \frac{1}{\beta} \sum_n^{\;\;\;\;\;\;\; \prime}
\int_0^{\pi/2} \!\!\!\!\!\!\! \mr{d}\phi
\frac{2N_0 V_d \Dd\cos^2\phi}{\sqrt{\wt^2+\Dst^2+\Dd^2\cos^2\phi}}
\end{eqnarray}
which leads indeed to expressions~(\ref{gap1_sid})-(\ref{gap2_sid}).

The $1/\w$ large frequency behavior in the Matsubara sum is due to taking the
infinite bandwidth limit, and a frequency cutoff $\w_c$ must be introduced. In
practice we have not employed a hard cutoff, as this leads to a first-order
jump of the order parameter at the critical temperature, but rather a
smooth cutoff function, $\sum_n^{\;\; \prime} \rightarrow
\sum_n f(\w_n)$, with:
\begin{equation}
\nonumber
f(\w) =
\left\{
\begin{array}{ll}
1                    & \mbox{if $\w<\w_c$} \\
e^{-2(\w-\w_c)/\w_c} & \mbox{if $\w>\w_c$} \\
\end{array}
\right.
\,.
\end{equation}

Finally, let us determine the location of the critical points in the clean limit
\cite{ChubuJyont}.
Subtracting twice~(\ref{GAP2_sid}) from~(\ref{GAP1_sid}) at zero temperature
gives:
\begin{equation}
\nonumber
V_s^{-1} - 2V_d^{-1} =
2N_0 \int_0^{\pi/2} \!\!\!\!\!\!\! \mr{d}\phi
\int_{-\w_c}^{\w_c} \frac{\mr{d}\w}{2\pi} \frac{\cos(2\phi)}
{\sqrt{\w^2+\Ds^2+\Dd^2\cos^2\phi}}
\end{equation}
Performing the $\w$-integral in the large $\w_c$ limit, we find:
\begin{eqnarray}
\nonumber
V_s^{-1} - 2V_d^{-1} &=&
- 2N_0 \int_0^{\pi/2} \!\!\!\!\!\!\! \mr{d}\phi \;
\cos(2\phi) \; \log\left(\alpha^2+\cos^2\phi\right)\\
&=& \frac{N_0}{2}\left(-1-2\alpha^2+2\alpha\sqrt{\alpha^2+1}\right)
\end{eqnarray}
where $\alpha \equiv \Ds/\Dd$. The instability criterion for the $s$-wave phase
(resp. $\dd$-wave) is given by $\alpha=0$ (resp. $\alpha=\infty$), leading to
the mixed region~(\ref{window_sid}).

\subsection{$\dxy+i\dd$ superconductor}

The derivation of the gap equation is very similar to the previous case, and we
only quote the intermediate result:
\begin{eqnarray}
\nonumber
\Dxy &=& - \frac{1}{\beta} \sum_n^{\;\;\;\;\;\;\; \prime}
\int_0^{\pi/2} \!\!\!\!\!\!\! \mr{d}\phi
\frac{2N_0 V_{xy} \Dxy\sin^2\phi}{\sqrt{\wt^2+\Dxy^2\sin^2\phi+\Dd^2\cos^2\phi}} \,, \\
\label{GAP1_did}
\\
\nonumber
\Dd &=& - \frac{1}{\beta} \sum_n^{\;\;\;\;\;\;\; \prime}
\int_0^{\pi/2} \!\!\!\!\!\!\! \mr{d}\phi
\frac{2N_0 V_d \Dd\cos^2\phi}{\sqrt{\wt^2+\Dxy^2\sin^2\phi+\Dd^2\cos^2\phi}}
\\
\label{GAP2_did}
\end{eqnarray}
which gives expressions~(\ref{gap1_did})-(\ref{gap2_did}) after some manipulations.

We also find the location of the critical points in the clean limit.
Subtracting~(\ref{GAP2_did}) from~(\ref{GAP1_did}) at zero temperature and
performing the $\w$-integral at large $\w_c$ leads to:
\begin{eqnarray}
\nonumber
V_{xy}^{-1} - V_d^{-1} \!\!\!&=&\!\!\!
- 2N_0 \!\!\int_0^{\pi/2} \!\!\!\!\!\!\! \mr{d}\phi
\cos(2\phi) \log\left(\frac{\alpha^2}{1-\alpha^2}+\cos^2\phi\right)\\
&=& \frac{N_0}{2} \frac{\alpha-1}{\alpha+1}
\end{eqnarray}
where $\alpha \equiv \Dxy/\Dd$ now. The instability criterion for the $\dxy$-wave
phase (resp. $\dd$-wave) is given by $\alpha=0$ (resp. $\alpha=\infty$), leading
to the mixed region~(\ref{window_did}).

\section{Dirty superconductor with bond order}
\label{app2}

\subsection{Gap equation}

We give here some details on the derivation of the gap equation for the
bond-ordered superconductor.
Due to the broken lattice symmetry, the resulting
$2\times1$ unit cell leads us to introduce two fermion species on each sublattice,
$c_{Ij\s}^\dagger$ and $d_{Ij\s}^\dagger$, and we can directly express the
Hamiltonian~(\ref{HMF}) in real space in those variables. It is then convenient
to go to Fourier space, and we set
\begin{eqnarray}
c_{Ij\s}^\dagger & = & \sum_{K_x} \sum_{k_y} c_{Kxk_y\s}^\dagger \; e^{i(K_x I + k_y j)} \,, \\
d_{Ij\s}^\dagger & = & \sum_{K_x} \sum_{k_y} d_{Kxk_y\s}^\dagger \; e^{i(K_x I + k_y j)} \,,
\end{eqnarray}
where $K_x = 2\pi/(L/2),\ldots,2\pi$ and $k_y = 2\pi/L,\ldots,2\pi$ take
respectively $L/2$ and $L$ values ($L\times L$ is the total number of sites).
Inserting these expression in the mean-field Hamiltonian gives readily the final
matrix expression~(\ref{Hmatrix}).

We can also write the gap equation~(\ref{DeltaRR}) in terms of these degrees of
freedom:
\begin{eqnarray}
\nonumber
\Delta_1 & = & -\frac{J}{\beta} \sum_{\w_n K_x k_y} \sum_\s (-1)^\s
\big<d_{-K_x-k_y-\s}^{\phantom{\dagger}} c_{K_xk_y\s}^{\phantom{\dagger}} \big> \\
\nonumber
\Delta_2 & = & -\frac{J}{\beta} \sum_{\w_n K_x k_y} \sum_\s (-1)^\s e^{iK_x}
\big<d_{-K_x-k_y-\s}^{\phantom{\dagger}} c_{K_xk_y\s}^{\phantom{\dagger}} \big> \\
\nonumber
\Delta_3 & = & -\frac{J}{\beta} \sum_{\w_n K_x k_y} \sum_\s (-1)^\s e^{ik_y}
\big<c_{-K_x-k_y-\s}^{\phantom{\dagger}} c_{K_xk_y\s}^{\phantom{\dagger}} \big>
\end{eqnarray}
Introducing the propagators $g_{mn} \equiv \big< \Psi_n^\dagger
\Psi_m^{\phantom{\dagger}} \big>$, we can write the gap equations as:
\begin{eqnarray}
\label{D1}
\Delta_1 & = & -\frac{J}{\beta} \sum_{\w_n K_x k_y}
\big[g_{14} + g_{32}\big] \\
\label{D2}
\Delta_2 & = & -\frac{J}{\beta} \sum_{\w_n K_x k_y}
\big[e^{iK_x}g_{14} + e^{-iK_x}g_{32}\big] \\
\label{D3}
\Delta_3 & = & -\frac{J}{\beta} \sum_{\w_n K_x k_y}
2\cos k_y \; g_{12}
\end{eqnarray}

\subsection{T-matrix}

We prove here the formula~(\ref{Tmatrix}) for the T-matrix of a bond-ordered
superconductor. We consider a single impurity at site ${\bf R=0}$, and start
with the exact Dyson's equation $\big[i\w - H_{MF} - v \sum_\s f_{{\bf 0}\s}^\dagger
f_{{\bf 0} \s}^{\phantom{\dagger}}\big] g = 1$, which we want to sandwich between two Bloch
states $\big|k,n\big>$, where $k=(K_x,k_y)$ and $n=1,\ldots,4$ is the Nambu
index. Introducing the propagator $g_{mn}(k,k') = \big<k,m\big|g\big|k',n\big>$,
we find in matrix (Nambu) notation:
\begin{eqnarray}
\big[i\w - \hat{h}_k \big] \hat{g}(k,k') & = &
\delta_{kk'} + v \sum_{k''} \hat{D}
\hat{g}(k'',k') \\
\hat{D} & \equiv & \left[
\begin{array}{cccc}
1 & 0 & 0 & 0 \\
0 & -1 & 0 & 0 \\
0 & 0 & 0 & 0 \\
0 & 0 & 0 & 0
\end{array}
\right]
\end{eqnarray}
Performing the matrix algebra, we can solve the previous equation, and we find
the local propagator:
\begin{eqnarray}
\hat{G}(k,k') & = & \frac{\delta_{kk'}}{i\w - \hat{h}_k} +
\frac{1}{i\w - \hat{h}_k} \hat{T}(i\w)
\frac{1}{i\w - \hat{h}_{k'}} \\
\hat{T}(i\w) & = & v \hat{D} \left(
1-v\sum_k \frac{1}{i\w-\hat{h}_k} \hat{D} \right)^{-1} \\
\nonumber
& = & \frac{1}{\mr{Det}}
\left[
\begin{array}{cccc}
-v^{-1}-G_{22} & G_{21} & 0 & 0 \\
G_{12} & v^{-1}+G_{11} & 0 & 0 \\
0 & 0 & 0 & 0 \\
0 & 0 & 0 & 0
\end{array}
\right]
\end{eqnarray}
where we have defined:
\begin{eqnarray}
\nonumber
\mr{Det} & \equiv & G_{11}G_{22} - G_{12}G_{21}-v^{-1}\big(G_{11}+G_{22}\big)-v^{-2} \\
\end{eqnarray}
Taking into account the contribution from site $\R = {\bf e}_x$, and in the
unitary limit $v=\infty$, we find the final expression~(\ref{Tmatrix}).

\bibliographystyle{apsrev}

\newcommand{\npb}{Nucl. Phys.}\newcommand{\adv}{Adv.
  Phys.}\newcommand{\epl}{Europhys. Lett.}

\end{document}